\documentclass[twocolumn,showpacs,preprintnumbers,amsmath,amssymb,superscriptaddress]{revtex4}

\usepackage{graphicx}
\usepackage{dcolumn}
\usepackage{bm}

\newcommand{\bx}{\text{box}}

\newcommand{\hc}{}

\begin{document}

\preprint{}

\title{Cooling and aggregation in wet granulates}

\author{Stephan Ulrich}
\email{ulrich@theorie.physik.uni-goettingen.de}
\affiliation{Universit\"at G\"ottingen, Institute of Theoretical Physics, Germany}%

\author{Timo Aspelmeier}%
\affiliation{
Max-Planck-Institut f\"ur Dynamik und Selbstorganisation, Dept.~Dynamics of Complex Fluids, G\"ottingen, Germany
}%

\author{Klaus R\"oller}
\affiliation{
Max-Planck-Institut f\"ur Dynamik und Selbstorganisation, Dept.~Dynamics of Complex Fluids, G\"ottingen, Germany
}%

\author{Axel Fingerle}
\affiliation{
Max-Planck-Institut f\"ur Dynamik und Selbstorganisation, Dept.~Dynamics of Complex Fluids, G\"ottingen, Germany
}%

\author{Stephan Herminghaus}
\affiliation{
Max-Planck-Institut f\"ur Dynamik und Selbstorganisation, Dept.~Dynamics of Complex Fluids, G\"ottingen, Germany
}%

\author{Annette Zippelius}%
\affiliation{Universit\"at G\"ottingen, Institute of Theoretical Physics, Germany}%
\affiliation{
Max-Planck-Institut f\"ur Dynamik und Selbstorganisation, Dept.~Dynamics of Complex Fluids, G\"ottingen, Germany
}%

\date{\today}

\begin{abstract}
  Wet granular materials are characterized by a defined bond energy in
  their particle interaction such that breaking a bond implies an
  irreversible loss of a fixed amount of energy. Associated with the
  bond energy is a nonequilibrium transition, setting in as the
  granular temperature falls below the bond energy. The subsequent
  aggregation of particles into clusters is shown to be a self-similar
  growth process with a cluster size distribution that obeys
  scaling. In the early phase of aggregation the clusters are fractals
  with $D_\text{f}=2$, for later times we observe gelation. We use simple
  scaling arguments to derive the temperature decay in the early and
  late stages of cooling and verify our results with event-driven
  simulations.
\end{abstract}

\pacs{45.70.-n, 47.57.-s, 61.43.Hv}
\maketitle

Granular systems have encountered strongly increasing interest within
recent years \cite{Jaeger96,D00} as they are both close to industrial
applications and relevant as model systems for collective systems far
from thermal equilibrium. Most research has focused on \emph{dry}
granulates resulting in rich non-equilibrium phenomena
\cite{PoeschelBuch}. The dramatic change of mechanical properties due
to some wetting liquid additive is apparent, when we compare the
fluid-like state of dry sand flowing through an hourglass with the
shapeable plastic state of wet sand, which is suitable for molding a
sandcastle. This change in bulk properties can be attributed to the
differences in the underlying particle
interactions \cite{Willett00,HerminghausAdv2005}. Whereas in collisions of dry
particles, a certain fraction of the initial kinetic energy is
dissipated into the atomic degrees of freedom of the particles, in the
wet case, the interaction is mainly due to the interfacial forces
exerted by liquid capillary bridges which form between adjacent
particles and dissipate energy upon rupture.

More recently, the dynamics of wet granular media has been addressed
in several
studies \cite{Thornton95,Lian98,Huang05,HerminghausAdv2005,Zaburdaev06,Fingerle08},
focusing on nonequilibrium phase
transitions \cite{Fingerle08},
agglomeration \cite{Thornton95,Lian98}, shear flow \cite{Huang05} and
cooling in one dimension \cite{Zaburdaev06}.  A particularly important
aspect of free cooling in cohesive gases is the structure of the
emerging clusters. It pertains to the formation of dust filaments and
the microscopic mechanisms of cloud formation as well as to the size
distribution and impact probability of planetesimals in accretion
discs. Structure formation in wet granulates during free cooling has
hardly been studied yet and is the central theme of our paper. We
suggest a very simple model for the interaction of two wet grains,
which only takes into account the essential features of a capillary
bridge: hysteresis and dissipation with a well defined energy loss.
Cooling is controlled by the probability for a bridge to rupture and
hence logarithmically slow in the long time limit, when a percolating
structure has been formed. For smaller times the structure is
characterized by coexisting fractal clusters of all sizes, whose size
distribution is shown to scale.

{\it Model---}Consider a gas of $N$ hard spheres of diameter $d$ in a
volume $V=L^3$. Each particle is covered with a thin liquid film. When
two particles collide, a liquid bridge forms between the two spheres
and induces an attractive force by virtue of the surface tension of
the liquid \cite{Willett00,HerminghausAdv2005}. As the spheres
withdraw from each other and their separation $s$ increases, the
bridge continues to exert a capillary force, $F(s)$, up to some
critical distance, $d_c>d$, where the bridge ruptures and liberates
the particles. Thus there is a hysteretic interaction between the
grains. Each rupture of a liquid bridge gives rise to dissipation of a
fixed amount of energy, $\Delta
E=\int_0^{d_c}F(s)ds$. Note that during each
collision the total momentum of the impact partners is
conserved. 
If the relative velocity of the impacting
particles is insufficient to rupture the liquid bridge,
$v<v_{\text{crit}}=\sqrt{2\Delta E/\mu}$ (where $\mu$ is
the reduced mass), they form a bounded state 
and continue their motion with the center
of mass velocity. 

We have simulated the above model with an event driven algorithm,
assuming the force between two colliding particles to be zero, except
when they touch -- such that their relative distance is equal to $d$
-- or recede beyond the critical distance $d_c$ for bridge
rupture. When the particles touch, they are elastically reflected
(complete restitution).  When the colliding particle move apart and
cross the critical distance $d_c$, the energy $\Delta E$ is taken out
of the motion of the impact partners - provided
$v>v_{\text{crit}}$. Otherwise the particles are elastically reflected
and remain in the bounded state. In a realistic model of wetted grains
(thin film), the capillary film is infinitesimally thin, so that it is
justified to assume that a bridge is only formed, if the particles
actually touch. This implies that energy is removed at $d_c$ only, if
the grains have touched each other before. In an alternative model
(thick film) one assumes that a bridge is formed, if the particles
approach closer than $d_c$. The difference between the two models is
insignificant for our results and is
discussed in more detail in \cite{InPreparation}. We use dimensionless
units such that $\Delta E = 1$, particle mass $m = 1$ and particle
diameter $d = 4$. The bond-breaking distance is chosen as $d_c = 1.07
d$ and volume fraction is varied from $\phi \approx 0.06\%$ up to
$15.6\%$.  We use periodic boundary conditions in the $x$- and
$y$-direction and hard walls in $z$-direction.

\emph{Cooling---}We define the granular temperature $T = \frac{1}{3N}
\sum_{i=1}^N m {\bf v}_i^2$ and investigate its decay in time from a
given initial value $T_0\gg\Delta E$. In each collision an amount of
energy $\Delta E$ is lost with probability $P_{bb}$, the probability
to break a capillary bridge. Hence the average loss of energy per time
is given by
\begin{equation}
  \frac{3}{2}\frac{dT}{dt} = - \frac{1}{2} \cdot f_\text{coll}\cdot  \Delta E \cdot P_{bb}\, . 
\label{eq:dE/dt}
\end{equation}
where $f_\text{coll}$ denotes the collision frequency. In the early
phase of cooling $T \gg \Delta E$, so that every collision also leads
to a rupture of the created capillary bond and $P_{bb}\approx 1$.  
The collision frequency, $f_\text{coll} = 4 g(d) \sigma n \sqrt{{T}/{\pi m}}$
is well established for a dilute system, with the particle density
$n=\frac{N}{V}$, the scattering cross section $\sigma$ and the pair
correlation function at contact $g(d)$. For the latter we use the
Carnahan-Starling approximation~\cite{Carnahan69}.
 The simplified cooling equation
$\frac{dT}{dt} \sim - \sqrt{T} $ is easily solved by
\begin{equation}
 T(t) = \left\{ \begin{array}{lll}
                 T_0\; (1-t/t_0)^2 \quad & \text{for } & t \leq t_0 \\
                 0                                  & \text{for } & t > t_0
                \end{array}
        \right. \label{eq:modelT(t)}
\end{equation}
with a characteristic time scale
\begin{equation}
 t_0 = \frac{3 \sqrt{\pi m T_0}}{2 g(d) \sigma n \Delta E} \,. \label{eq:t0}
\end{equation}

\begin{figure}[h]
 \includegraphics[width=.49\textwidth]{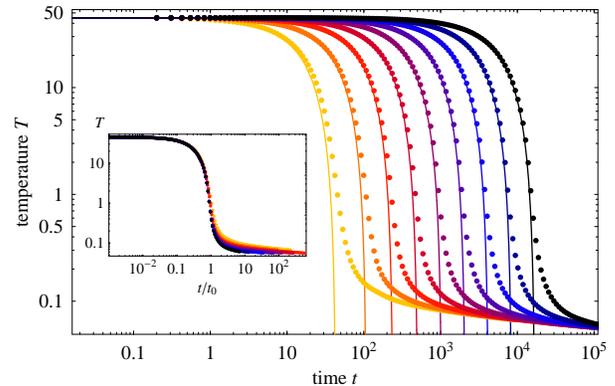}
 \centering
 \caption{(color online) Decay of the granular temperature $T$ for
   volume fractions (from left to right) $\phi = 15.6\%, 7.81\%,
   3.90\%,1.95\%, 0.98\%, 0.49\%, 0.24\%,0.12\%,$ $0.061\%$;number of
   particles $N=262144$ is fixed; comparison of theory (full line) and
   simulation ($\bullet$) inset: $T$ versus scaled time
   $t/t_0$.} 
 \label{fig:Temp}
\end{figure}

In Fig.~\ref{fig:Temp} the evolution of the granular temperature $T$
from the simulation (dots) is compared to the analytic result (full
lines) for different volume fractions $\phi$. 
The main dependence on $\phi$ is contained in the timescale
$t_0$. Hence the different curves can be approximately collapsed onto a
single curve, if $T$ is plotted versus scaled time $t/t_0$ (see inset).

The agreement between theory and simulation is quite good up to $t_0$, when
the assumption contained in eq.~(\ref{eq:dE/dt}) that \emph{every}
collision causes an energy loss $\Delta E$  breaks down. Yet
the timescale $t_0$ has a clear physical relevance. At $t_0$ the
temperature is comparable to the bond-breaking energy $\Delta E$ and
for $t>t_0$ persistent clusters will form with important consequences
for the cooling dynamics. Thus the timescale $t_0$ separates two
regimes with qualitatively different {\it structural} and {\it
  dynamic} properties.

\begin{figure}[t]
\includegraphics[width=.4998\textwidth]{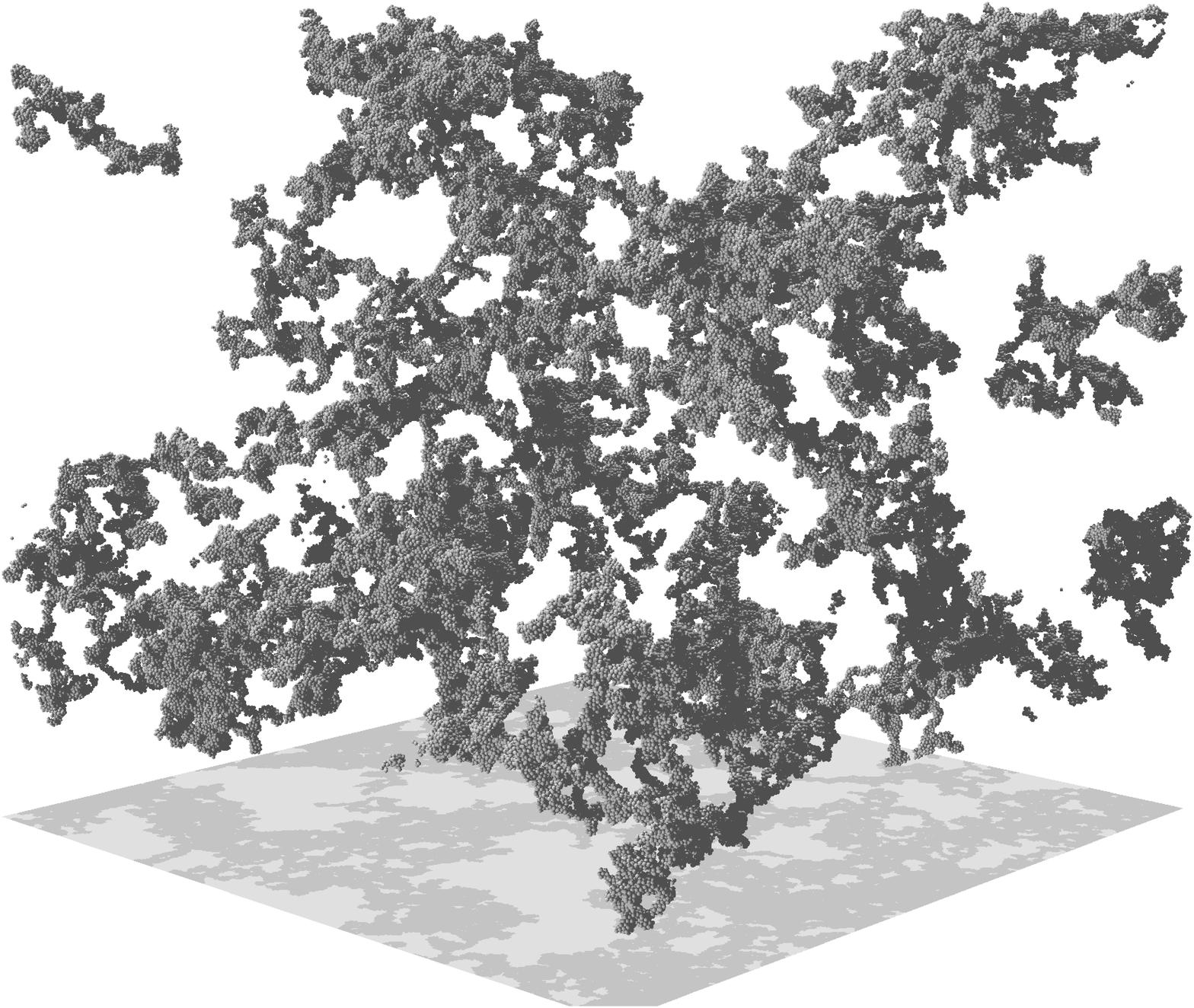}
\includegraphics[bb=600 530 620 531,width=.01\textwidth]{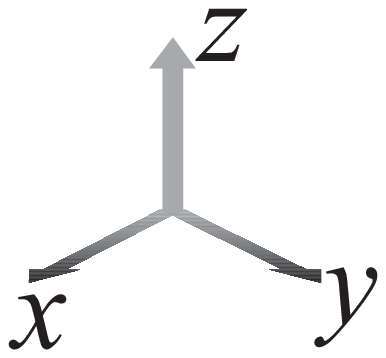}
 \centering
 \caption{Snapshot of the system with $\phi = 0.98\%$,
   $N=262144$ and $T \approx 0.06 \Delta E$ taken at time $t \approx
   36 t_0$; only particles in the spanning cluster are shown. A rotating version of this figure gives a better impression of the three-dimensional structure (download at \cite{RotatingCluster}).}
 \label{fig:SystemSnapshot}
\end{figure}

{\it Aggregation---}For $t> t_0$ the system starts to form aggregates,
which grow in a self-similar process, reminiscent of ballistic
cluster-cluster aggregation \cite{Jullien87}. For even longer times we
observe a spanning or percolating cluster for all finite densities and
ultimately all particles and clusters have merged into a single
cluster. In Fig.~\ref{fig:SystemSnapshot} we show such a cluster which
has attracted $99.7\%$ of all particles.

\begin{figure}[h]
 \includegraphics[width=.49\textwidth]{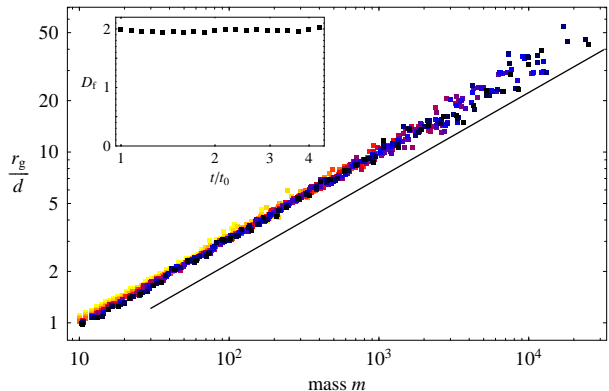}
 \centering
 \caption{(color online) Radius of gyration as a function of cluster
   size for a system of 262144 particles at volume fraction
   $\phi=1.95\%$; different colors correspond to simulation times
   between $t_0$ (yellow) and $4t_0$ (black); the solid line has a slope of $1/2$; inset: fractal
   dimension for each snapshot separately.}
 \label{radius_gyration}
\end{figure}

The structure of the aggregates in the {\it early} stage of
aggregation, sometimes called flocculation regime, is characterized by
the fractal dimension $D_\text{f}$.  It is easily determined from the
radius of gyration, which is expected to scale with the cluster mass
$m$ like $r_\text{g} \sim m^{1/D_\text{f}}$, defining $D_\text{f}$.
In Fig.~\ref{radius_gyration} we show the radius of gyration for a
system of $262144$ particles at volume fraction $\phi=1.95\%$. Several
snapshots of the ensemble of growing clusters have been taken at times
$t_0<t<4t_0$ (nonpercolating regime). The data scale well according to
the above relation, some scatter is observed for the largest masses,
corresponding to times close to the percolation transition. Since our
clusters are not completely static, we plot the fractal dimension for each snapshot separately in the inset of Fig.~\ref{radius_gyration}. The variation is very small so that we conclude that $D_\text{f}$ does not depend on time.

\begin{figure}[h]
 \includegraphics[width=.49\textwidth]{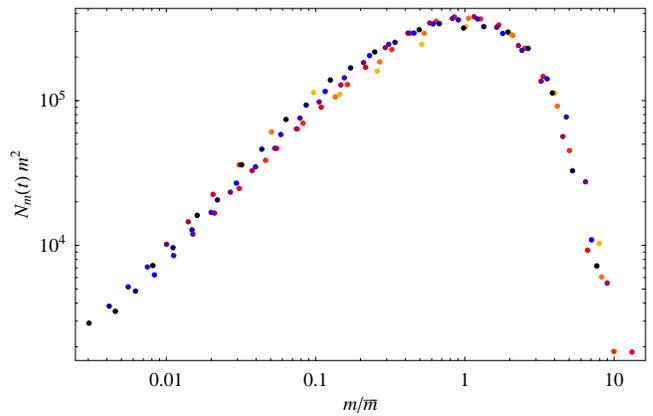}
 \centering
 \caption{(color online) Rescaled cluster size distribution $f(m /
   \bar{m}) = N_m(t) \cdot m^{2}$ versus scaled cluster mass $m /
   \bar{m}$ for $\phi=1.96\%$; color coding refers to different
   snapshots.}
 \label{fig:ClusterSizeDistLetter}
\end{figure}

All information about the connectivity of the clusters is contained in
the {\it cluster size distribution}, $N_m(t)$, the number of clusters of size
$m$ at time $t$. For aggregating systems the mass distribution is
expected to evolve towards a self-preserving scaling
form \cite[e.g.][]{meakin91}, independent of the initial distribution:
\begin{equation}
 N_m(t) = m^{-\theta} f\big(m / \bar{m}(t)\big) \, ,\label{eq:ClusterScaling}
\end{equation}
where the time dependence is only contained in the mean cluster mass
$\bar{m}(t)=\sum_{m=1}^{\infty} m^2 N_m(t)/N$.  Mass conservation
requires $\theta = 2$ \cite{meakin91}.  We plot in
Fig.~\ref{fig:ClusterSizeDistLetter} the scaling function $f(m /
\bar{m})= N_m(t) m^2 $ for the data from eight snapshots taken at
times $t$ between $t_0$ and $2t_0$. The data are seen to scale well in
this regime, in which the mean cluster mass increases roughly by a factor
of $30$.

The {\it asymptotic} cluster, which contains all the particles, is
compact on large scales. However, its structure at smaller length
scale may well be fractal. To investigate the Hausdorff dimension of
the largest cluster at {\it intermediate} length scales, we use the
box counting algorithm \cite{grassberger83,hentschel83} and determine
the number of boxes $N_\bx$ of edge length $L_\bx$, needed to cover
the whole cluster \footnote{For the evaluation we use a slightly different algorithm, where particles are assumed to be point-like. A difference, however, would only be observable in the irrelevant regime $L_\bx < d$.}. On length scales much smaller than the particle
diameter, $L_\bx \ll d\hc$, the system obviously behaves
three-dimensionally and $N_\bx = \phi (L/L_\bx)^3$. Since our
system is finite and contains a system-spanning cluster, the scaling
behavior on large length scales should also be three-dimensional:
$N_\bx = (L/L_\bx)^3$ for $L_\bx \approx L$. Only in the
intermediate regime, whose size is proportional to $|\log \phi|$, can
we expect to find a nontrivial $D_\text{f}$ such that $N_\bx \sim L_\bx^{-D_\text{f}}$.

\begin{figure}[h]
 \includegraphics[width=.49\textwidth]{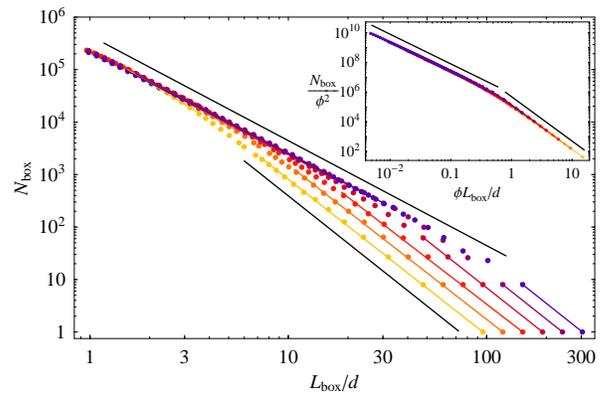}
  \centering
  \caption{(color online) Result of the box counting algorithm for $N
    = 262144$, $t \approx 45 t_0$ and varying volume fraction $\phi =
    15.6\%, 7.81\%, 3.90\%, 1.95\%, 0.98\%, 0.49\%$ (from left to
    right); the lines along the data points are fits; the outer lines
    have slope $-2$ (top) and $-3$ (bottom); inset: scaling plot of
    $N_\bx/\phi^2$ versus $L_\bx \phi/d$.}
 \label{fig:HausdorffScaleDensity}
\end{figure}

Results of the box counting algorithm for different densities (all in
the percolating regime) are presented in
Fig.~\ref{fig:HausdorffScaleDensity}. One can clearly distinguish two
scaling regimes: at intermediate length scales a nontrivial slope of
$-2$ is observed, before the crossover to the expected value of $-3$
on the largest scales takes place. The size of the nontrivial regime
shrinks with increasing density according to $|\log\phi|$. Thus the
crossover length scales as $L_\bx/d\propto 1/\phi$ and the number of
boxes at this length scales is $N_\bx=(L/L_\bx)^3\propto
\phi^2$. Hence if we plot $N_\bx /\phi^2$ versus $L_\bx \phi/d$, the
data should collapse onto a single universal curve. This is indeed the
case, as can be seen in the inset of
Fig.~\ref{fig:HausdorffScaleDensity}.



\begin{figure}[h]
 \includegraphics[width=.49\textwidth]{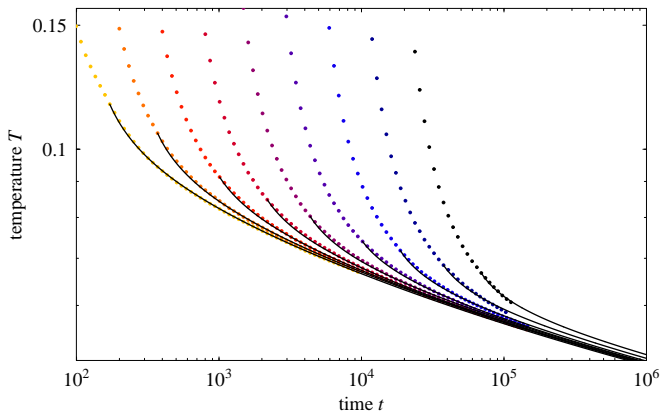}
  \centering
  \caption{(color online) Asymptotic time dependence for several
    volume fractions as in Fig.(\ref{fig:Temp}); data (dots) in
    comparison to the analytical results (lines)}
 \label{fig:asymptotic_time}
\end{figure}
{\it Long time asymptotics---}In the late stage of cooling the probability
$P_{bb}$ to break a bond is given by the probability to find a kinetic
energy larger than $\Delta E$:
\begin{equation}
P_{bb}=\int d^3v \,\theta(mv^2/2-\Delta E)\, w({\bf v}). 
\end{equation}
We approximate the velocity distribution, $ w({\bf v})$, by a Maxwellian
of width $T(t)$
and evaluate the above integral in the limit $T(t)/\Delta E \to
0$. The probability to break a bond becomes exponentially small in that
limit
\begin{equation}
P_{bb}= \left(\frac{4\Delta E}{\pi T}\right)^{1/2}\, e^{-\Delta E/T}
\end{equation}
dramatically slowing down the decay of the kinetic energy according to
Eq.(\ref{eq:dE/dt}): 
$\frac{dT}{dt}\sim \exp{(-\Delta E/T)} $, predicting a logarithmic
decay of $T$ for asymptotically large times.
The full solution of the above
equation is in good agreement with the simulation data as shown in
Fig.~\ref{fig:asymptotic_time}. The proportionality constant in the
rate equation is fitted, because the collision frequency is not known
in this highly clustered state.


{\it Conclusions---}The characteristic energy loss $\Delta E$ of wet
granulates gives rise to a well defined dynamic transition at time
$t_0$, when the kinetic energy $T$ is equal to $\Delta E$. For $t<t_0$
very few collisions give rise to bound pairs of particles and cooling
proceeds according to $T(t) = T_0\; (1-t/t_0)^2 $. For $t>t_0$ the
particles aggregate into clusters with a flocculation regime in the
early stages of aggregation and subsequent percolation. Cluster growth
is a self similar process with a cluster size distribution that obeys
scaling. The asymptotic time decay is logarithmically slow, because
bond-breaking becomes a very unlikely process.

\begin{acknowledgments}
  We gratefully acknowledge financial support by the Deutsche
  Forschungsgemeinschaft (DFG) through Grant SFB 602/B6.
\end{acknowledgments}

\bibliographystyle{apsrev}
\bibliography{WetGranulates}

\end{document}